# Research in Global Software Engineering: A Systematic Snapshot

Bilal Raza, Stephen G. MacDonell, and Tony Clear

SERL, School of Computing & Mathematical Sciences, Auckland University of Technology,
Private Bag 92006, Auckland 1142, New Zealand
bilal.raza@aut.ac.nz, stephen.macdonell@aut.ac.nz, tony.clear@aut.ac.nz

**Abstract**

*This paper reports our extended analysis of the recent literature ad- dressing global software engineering (GSE), using a new Systematic Snapshot Mapping (SSM) technique. The primary purpose of this work is to understand what issues are being addressed and how research is being carried out in GSE – and comparatively, what work is not being conducted. We carried out the analysis in two stages. In the first stage we analyzed 275 papers published between January 2011 and June 2012, and in the second stage we augmented our analysis by considering a further 26 papers (from the 2013 International Conference on Global Software Engineering (ICGSE'13). Our results reveal that, currently, GSE studies are focused on management- and infrastructure-related factors, using principally evaluative research approaches. Most of the studies are con- ducted at the organizational level, mainly using methods such as interviews, surveys, field studies and case studies. The USA, India and China are major players in GSE, with USA-India collaborations being the most frequently studied, followed by USA-China. While a considerable number of GSE-related studies have been published since January 2011 they are currently quite narrowly focused, on exploratory research and explanatory theories, and the critical research paradigm has been untouched. An absence of formulative research, experimentation and simulation, and a related focus on evaluative approaches, all suggest that existing tools, methods and approaches from related fields are being tested in the GSE context, even though these may not be inherently applicable to the additional scale and complexity of GSE.*

**Keywords:** Global Software Engineering (GSE), Distributed Software Development, Classification, Systematic Mapping.

## 1. INTRODUCTION

Interest in software development carried out by globally distributed, culturally and/or temporally diverse teams arose with the advent of outsourcing in the last two decades, and it continues to increase [1]. Its importance has led to the emergence of the specific area of research and practice referred to as global software engineering (GSE) [1]. GSE is itself a growing field as is clearly evident in the diversity of locations involved and the rapidly increasing number of published studies into GSE-related issues. As the number of such studies increases it becomes important to periodically summarize the work and provide overviews of the results [2] as a means of reflection on what work is being done and what gaps might exist.

In this paper we investigate the breadth of topics that have been covered by GSE studies over a short timeframe, using a variant of the systematic mapping (SM) method that we refer to as a *systematic snapshot*. This approach establishes a specific baseline state that could be further extended in a backward or forward direction to analyse changes over time. The systematic mapping (SM) method has been widely used in medical research [2] and was first adopted in software engineering research by Bailey et al. [3]. A SM aims to provide a high-level view of the relevant research literature by classifying the work according to a series of defined categories and visualizing the status of a particular field of research [2][4]. This technique has been used recently in the GSE field [4][5][6][7][8]. In these studies specific aspects of GSE research were categorized (using guidelines presented in [2] and [9]). These investigations considered between 24 and 91 primary studies, published up to the year 2010. The aspects of GSE analyzed in these studies were software configuration management, awareness support, agile practices, project management, and tools in GSE. All five studies therefore classified the GSE literature from a relatively narrow perspective but covering a wide temporal range. They were published in well-known journals and conferences and provide valuable contributions to the body of GSE literature. In our study, we instead use a new variant of the systematic mapping process called Systematic Snapshot Mapping (SSM), briefly described in the next section, to classify the very recent global software engineering literature.

The rest of this paper is organized as follows: in Section 2 we describe our research approach in greater detail, and in Section 3 we present the findings of our analysis. In the subsequent Section 4 we briefly discuss validity threats. In Section 5 we conclude this paper and Section 6 conveys future work.

## 2. METHOD AND CONDUCT

The results presented in this paper derive from our classification of the current literature on GSE, using the Systematic Snapshot Mapping (SSM) method. In order to classify this literature we chose the time period between January 2011 and June 2012 and later extended it to include papers published in the Proceedings of ICGSE'13. This study followed guidelines presented by Petersen et al. [2] for carrying out systematic mapping studies. However, instead of narrowing down the topic and considering a large temporal period, we limited the time span and considered the full breadth of topics covered. This study was inspired by several prior classifications of SE and GSE literature including that of Glass et al. [10], but instead of following a random sampling technique to select papers (as in [10]) we used a systematic process. We employed a defined protocol for choosing search strings and executing them against relevant databases to cover the breadth of GSE-related studies.

We defined our categories at the outset of our work and chose various dimensions to present the results, mainly leveraging the prior classifications of Richardson et al. [11] and Glass et al. [10]. We present our results in the form of tables, bar graphs and network analysis graphs to provide visual representations of the data. We believe such a snapshot approach is especially useful in cases where a field is changing rapidly and where there is consequently rapid growth in the research literature. This new approach for carrying out systematic mapping also provides an opportunity to effectively build upon different researchers' work by using different temporal ranges.

### 2.1 Research Questions
The following research questions were established for this study:
- RQ1. What are the factors, levels and locations investigated in the recent GSE literature?
- RQ2. How is research being carried out in GSE in regard to methods and approaches?

### 2.2 Search Strategy
Our search strategy was designed to keep the topic general while addressing a short time period to provide an up-to-date overview of the research literature. Initial search keywords were selected from known GSE systematic literature reviews and mapping studies. These keywords were updated based upon various dry runs carried out on the Scopus database to ensure their effectiveness. In the initial run, a target was set to ensure at least those studies from which the keywords were taken were retrieved. In the second run, a random set of ten studies was selected from the Proceedings of the 2009-2011 ICGSE conferences, and the search strings were further refined to ensure that these sample studies were also retrieved.

Table 1 shows the final list of keywords used to cover as many variations of the same term as possible. We intentionally adopted many keywords having low precision but high recall [12] and subsequently complemented our analysis by including all the papers published in ICGSE'13.

**Table 1.** List of keywords used as search strings.

"global software engineering" OR "global software development" OR "distributed software engineering" OR "distributed software development" OR "offshore software development" OR "offshore software engineering" OR "distributed team" OR "global team" OR "offshore insourcing" OR "geographically distributed teams" OR "global software" OR {("software" OR "information system" OR "computer" OR "information technology") AND ("virtual team" OR "dispersed team" OR "far shore" OR "offsite" OR "offshore outsource" OR "outsourced"}}

### 2.3 Data Sources and Retrieval
We searched across multiple data sources to retrieve as many potentially relevant studies as possible. SCOPUS, IEEE Xplore, the ACM Digital Library, SpringerLink and ScienceDirect were searched to complement results. Each database has limitations in terms of the number of keywords accepted at a specific instance; therefore, we had to break the search phrases to suit the particular database. The initial search and retrieval process was conducted in July 2012 and the date range was limited to January 2011 to June 2012. The search was carried out on metadata (title, abstract, keywords) and only peer-reviewed literature published in English was considered. In the first step, citations of retrieved studies were downloaded and duplicates were removed. Afterwards, the studies were then considered for the inclusion process.

### 2.4 Inclusion Process
The steps taken in the inclusion process to select primary studies are shown in Figure 1. After searching all of the databases 2020 studies were retrieved. The decision for further inclusion was based upon the first author's reading of the papers' titles or abstracts (resulting in 1125 studies). Duplicates were then removed, and a full text version of each remaining study was sought. For 12% of the papers (53 of the 437 remaining) the full text was not available to us, primarily because the papers were not published in well-known journals or conference proceedings. These studies were there-fore not considered for further analysis. The full text of the remaining 384 papers was then reviewed by the first author and a set of 275 studies was selected for inclusion in the SM analysis. Studies in the form of short papers, extended abstracts and position papers (only describing future work) were excluded. A number of studies, not related to the software engineering domain, had slipped through to this stage and upon cursory review of the full text were also excluded. At this stage, we also considered the papers of ICGSE'13 and included them in our final list for analysis.

### 2.5 Data Extraction and Synthesis
We followed generally accepted guidelines [2] to build our SM classification scheme. The included studies were therefore categorized according to various dimensions: research approach, research method, factors, level of analysis, sourcing phases and locations. In order to reduce threats to validity, regular meetings of the three authors were held to discuss issues and address misconceptions. In order to reduce bias effects the three researchers also conducted a sample classification together. At a later point a further sample of studies which were initially classified by the first author were verified by the senior researchers, discussions were held again and issues were addressed. It was established that the authors were in general agreement regarding the classification, based upon the sample results. The classification scheme utilized by Glass et al. [10] was used to characterize the research approach for our set of studies. We also considered the same source for the methodologies used in software engineering research. However, to better reflect the GSE perspective we also considered other methodologies [1][13]. Hence, we add- ed Computer Mediated Communication (CMC) analysis to cover studies that investigate artifacts such as chat-histories and emails. Although grouped together in prior studies, Observations and Interviews were considered separately, as many studies use them to complement other methods. Interviews are widely used as a sub-method in Case Studies and Observations are used in Ethnographies. However, we observed that these methods are being used in their own right and we therefore classified them separately. We included the method Data Analysis to signify studies that utilize data from Repositories, Incident Management Systems and Archives of previous projects. We used Proof of Concept for non-empirical studies in which entities were formulated but were only described by examples rather than any formal validation.

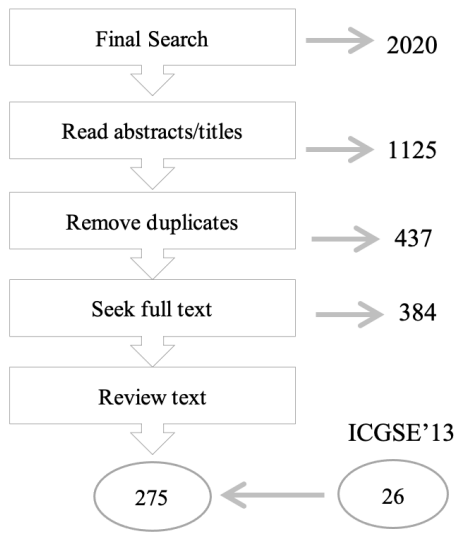

Fig. 1. Study inclusion process

## 3. FINDINGS

This section presents the results obtained based on the classifications of the data extracted from our final combined set of studies.

### 3.1 Findings for Factors

Richardson et al. [11] identified 25 GSE factors in an empirical study and grouped them in the four broad categories of Distance, Infrastructure, Management and Hu- man Factors. We used these categories to also characterize our identified studies. We added Learning/Training/Teaching, Competition and Performance to the Management category and Relationship to the Human Factors category. We also updated the latter category with Coordination/collaboration. Table 2 presents the results of this classification. The results clearly show that recent GSE studies are heavily focused on Management- and Infrastructure-related factors compared to Human- and Distance-related factors. Šmite et al. [1] presented a systematic review of empirical GSE research and also found that most of the studies were focused on management-related issues. Com- paring these results with the SWEBOK [14] knowledge areas (KAs), it was found that the standard lacks specific considerations for GSE. As a corollary, it was also found that KAs related to design, construction, testing and maintenance are not widely addressed in the recent GSE literature.

Table 2. Findings for GSE factors and their percentage.

| *Distance* | *16.4%* | Team Selection | 0.8% |
|---|---|---|---|
| Communication | 8.4% | Effective Partitioning | 4.6% |
| Language | 1.1% | Skills Management | 0.4% |
| Culture | 5.3% | Knowledge | 6.1% |
| Temporal issues | 1.6% | Visibility | 3.3% |
| *Human Factors* | *16.03%* | Reporting | 0.0% |
| Fear | 0.4% | Information | 1.1% |
| Motivation | 2.1% | Teamness | 5.5% |
| Trust | 2.7% | Learning/Training/tea | 4.2% |
| Cooperation | 1.6% | Competition | 0.6% |
| Coordination/col | 7.8% | Performance | 1.8% |
| Relationship | 1.2% | *Infrastructure* | *24.2%* |
| *Management* | *43.2%* | Process Management | 8.2% |
| True Cost | 1.7% | Tools | 9.1% |
| Project | 8.8% | Technical Support | 0.4% |
| Risk | 2.3% | Communication tools | 6.5% |
| Roles and responsibilities | 1.6% | | |

### 3.2 Findings for Research Approach

GSE presents a complex context that demands a more extensive repertoire of research methods and approaches than those currently prevailing [15]. Table 3 presents the findings of our classification of the research approaches used in current GSE-related studies. In terms of the three main categories, the dominant research approach is Evaluative, followed by Descriptive and then Formulative. This is in sharp contrast to the results reported in 2002 by Glass et al. [10] in which the order was Formulative, Descriptive and Evaluative. One of the main reasons for the present dominance of Evaluative research is the inclusion of new empirical methods such as CMC analysis, Interviews, Data Analysis and Observations. These results appear to be in contrast with the results of Šmite et al.'s systematic review [1] of GSE-related studies published between 2000 and 2008. They concluded that GSE-related studies are relatively small in number and immature and most of them focused on problem-oriented reports. Our current results show, however, that GSE publications have grown in quantity and quality and more studies have used evaluative approaches. Of note is that these evaluative approaches are mostly confined to previously formulated work. We interpret this to mean that existing methods, tools and so on from related fields, such as collocated software engineering (CSE), are being evaluated in the context of GSE. Given that GSE is fundamentally different from CSE[11], it seems likely that solutions formulated for CSE will need to be updated or enhanced for GSE. Entirely new

Table 3. Findings for research approach.

| Research Approach | Percentage | Research Approach | Percentage |
|---|---|---|---|
| *Descriptive* | *25.4%* | Evaluative-other | 12.1% |
| Descriptive-system | 7.4% | *Formulative* | *18.5%* |
| Review of literature | 9.8% | Formulative-framework | 5.2% |
| Descriptive-other | 8.1% | Formulative-guidelines/standards/approach (FG) | 1.6% |
| *Evaluative* | *56.1%* | Formulative-model | 5.9% |
| Evaluative-deductive | 17.6% | Formulative-process, method, algorithm | 2.3% |
| Evaluative-interpretive | 26.1% | Formulative-classification/taxonomy | 0.5% |
| Evaluative-critical | 0.2% | Formulative-concept | 2.7% |

solutions may also need to be identified and assessed in the GSE context.

Similarly, there is clear potential for critical research in this context particularly in light of the power structures that can exist between GSE 'partners', and the associated issues of trust, fear, cooperation and the like (as shown in Table 2). Criteria or principles for carrying out critical research are lacking generally in information systems (IS) [16]. Considering its importance, Myers and Klein [16] proposed a set of principles for conducting critical research – these principles could be considered in future investigations of human factors in GSE.

### 3.3 Findings for Research Methods

Figure 2 depicts the research methods used. The most dominant methods are Interview, Survey, Field Study and Case Study, indicating that most of the studies employed qualitative methods. These results are also in stark contrast to more general SE classifications [10] in which researchers used very few case or field studies. For studies in which multiple methods were used we assigned more than one research approach and method. Research methods in GSE are currently skewed towards exploratory research focusing on theories relating to 'Explanation' as described by Gregor [17]. These theories aim to provide explanation about what, how and why things happen and to promote greater understanding of phenomena. Thus, although GSE research has grown in terms of the number of studies being conducted, these studies are exploratory and/or explanatory in nature. It will be interesting to compare these results with future studies to determine whether work moves towards more predictive studies as the field matures.

### 3.4 Findings for Level of Analysis and Distribution of Studies

Figure 3 shows the level of analysis considered currently by GSE researchers. The dominant level of analysis was found to be Organizational followed by Inter-Organizational - combined together they are used in more than half the studies re- viewed. Fewer studies addressed group, individual and societal levels, a finding that coincides with the results of Glass et al. [10] in respect of SE studies. Table 4 presents the distribution of studies across various conferences, journals and workshops with frequency greater than one. (This limit was imposed due to space considerations and for ease of interpretation.) The majority of the selected studies was published in conference proceedings and drew on an industrial context.

### 3.5 Bubble Plot Analysis

The use of visual techniques in SM, such as bubble plots, has been recommended by Petersen et al. [2] and such techniques have been used to convey the results of mapping and classification studies[13][6]. Figure 4 presents the results of this study in the form of a bubble plot. We chose to represent three classification dimensions within it: Research approach is on the right X-axis, GSE-factors, grouped in their four major categories, are on the Y-axis, and level of analysis is on the left X-axis. The results clearly show that most of the recent studies are focused on using evaluative approaches around management and infrastructure factors and analyzed at the organizational levels. Stu- dies based upon specific groups, societies and individuals are limited. Organizational concerns have been at the forefront in terms of the level of analysis, leaving much scope for consideration of groups and individuals in future studies.

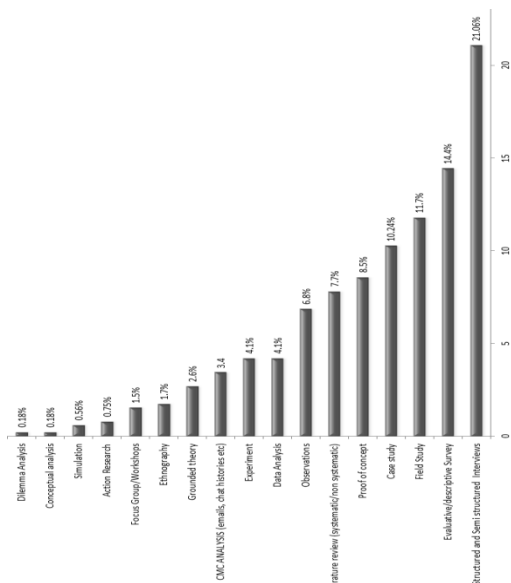

Fig. 2. Findings for research methods

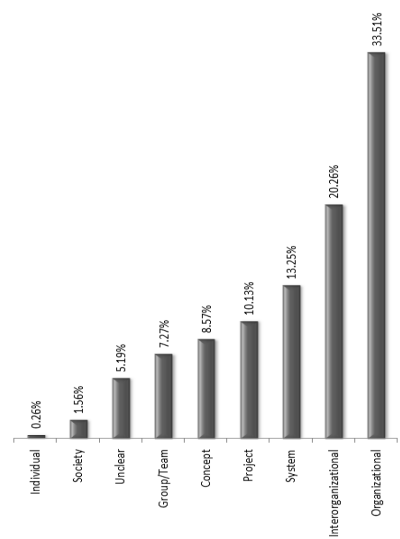

Fig. 3. Findings for level of analysis

Table 4. Distribution of studies across Journals, Conferences and Workshops.

| Journals | | | IEEE TEM | 2 | ISEC | 3 |
|---|---|---|---|---|---|---|
| | | | LNBIP | 2 | ICSSP | 3 |
| IST Journal | 8 | | J Grp Dec Negot | 2 | MySEC | 2 |
| JSEP | 7 | | **Conferences** | | EUROMICRO | 2 |
| J Softw. Maint. Evo. | 7 | | | | ICIS | 2 |
| IET Software | 6 | | ICGSE | 52 | CollaborateCom | 2 |
| J of E Markets | 4 | | HICSS | 15 | CTS | 2 |
| IEEE Software | 4 | | ICSE | 8 | PACIS | 2 |
| J Comm and Com Sc. | 3 | | CSCW | 8 | **Workshops** | |
| ISJ | 3 | | PROFES | 6 | CTGDSD | 13 |
| IJoPM | 2 | | CHI | 5 | ICGSE | 13 |
| JSW | 2 | | XP | 4 | CHASE | 7 |
| POM Journal | 2 | | ICIC | 3 | OTM | 3 |
| IS | 2 | | PICMET | 3 | Global Sourcing | 3 |

### 3.6 Location of GSE Projects and Inter-country Relationships

Figure 6 and Table 5 provide graphical and tabular representations of the locations involved in GSE projects. A few studies also mentioned regions rather than countries; we also considered them in our analysis. Figure 5 shows the results of our examination of inter-country networks. We used NodeXL, an extendable tool kit used for data analysis and visualizations [18]. Table 6 lists the pairwise relationships with frequency greater than one. (This constraint was imposed due to space limitations; however, all the relationships are shown in Figure 5.) It can be seen in Figure 5 and Table 6 that the most connected nodes are the USA and India. Some studies explicitly mentioned the collaborating locations whereas others only specified the locations involved with- out clearly stating which actively collaborated. For the latter studies, we assumed pairwise relationships between each location. For future studies we recommend that authors clearly state the nature of each party's involvement.

In Figure 6, countries and regions marked by darker shades are those most frequently involved in GSE. For ease of analysis we grouped these countries into six categories based upon the number of studies that cite their involvement. Not unexpectedly, the two countries reported as most frequently involved in global software projects are the USA and India. Countries including Germany, Finland, China, the UK, Australia and Brazil are ranked in the second group, closely followed by a group comprising Sweden, the Netherlands, Japan, Argentina, Spain, Canada and Switzer- land. In the next two categories lie the potentially upcoming and emerging countries of Russia, Eastern European countries such as Lithuania, Far Eastern countries including Malaysia and Indonesia, and the South/Central American countries of Chile and Mexico. These representations give some insight into the diversity of countries' involvement in GSE projects. Some of these regions are underrepresented but this does not necessarily mean that these locations are not involved in GSE; it could be that these regions have simply not been considered in recent studies.

Researchers rely on personal contacts in their national industries to validate their results. Our study also shows that the top seven locations of GSE authors are the USA, Finland, Germany, Spain, Brazil, India and Sweden. Apart from Spain, which is thirteenth, all six other countries are in the list of top ten locations involved in GSE projects. We also analyzed the inter-country collaboration of GSE researchers from different countries and found that researchers from European countries have mostly collaborated with other European-based researchers whereas researchers from the US have collaborated with European and Asian researchers.

### 3.7 Phases in Sourcing Relationships

Dibbern et al. [19] divided the sourcing process into two main stages: the decision stage, which is concerned with the 'What', 'Why' and 'Which' questions, and the implementation stage addressing 'How' and 'Outcome'. This covers the processes of deciding on and managing the sourcing resulting agreement, but leaves out the transition process. Butler et al. [20] divided this same process into three main phases, of Decision, Transition and Operation, based upon the timeline of a project. Butler et al. then [20] categorised 116 articles based upon the focus of attention of GSE projects and found that only 2 articles from the 116 were related to the transitional phase. This coincides with the results of this systematic snapshot mapping study in which we categorised 301 articles across various dimensions and found that only 19 were related to transition, further highliting that limited research has been directed towards this phase.

Table 5. Locations involved in GSE projects

| Country | # | Country | # | Country | # | Country | # | Country | # |
|---|---|---|---|---|---|---|---|---|---|
| US | 246 | Esp | 18 | Mys | 9 | Pan | 5 | Grc | 3 |
| Ind | 167 | Can | 16 | Mex | 9 | Aut | 4 | Twn | 3 |
| Deu | 61 | Che | 16 | Sen | 9 | Est | 4 | Rom | 2 |
| Fin | 56 | Ukr | 15 | SGP | 9 | Phl | 4 | Svk | 2 |
| Chn | 44 | Rus | 13 | Nzl | 9 | Tha | 4 | Tur | 2 |
| UK | 41 | Dnk | 13 | Hun | 8 | Vnm | 4 | Pak | 2 |
| Aus | 34 | Irl | 12 | Khm | 7 | Kor | 4 | Bgd | 1 |
| Bra | 32 | Ita | 11 | Fra | 7 | Pol | 4 | Zaf | 1 |
| Swe | 27 | Nor | 11 | Bel | 7 | Cri | 3 | Tun | 1 |
| Nld | 23 | Cze | 10 | Chl | 6 | Col | 3 | | |
| Jpn | 21 | Ltu | 10 | Hrv | 6 | Ecu | 3 | | |
| Arg | 19 | Isr | 9 | UAE | 5 | Egy | 3 | | |

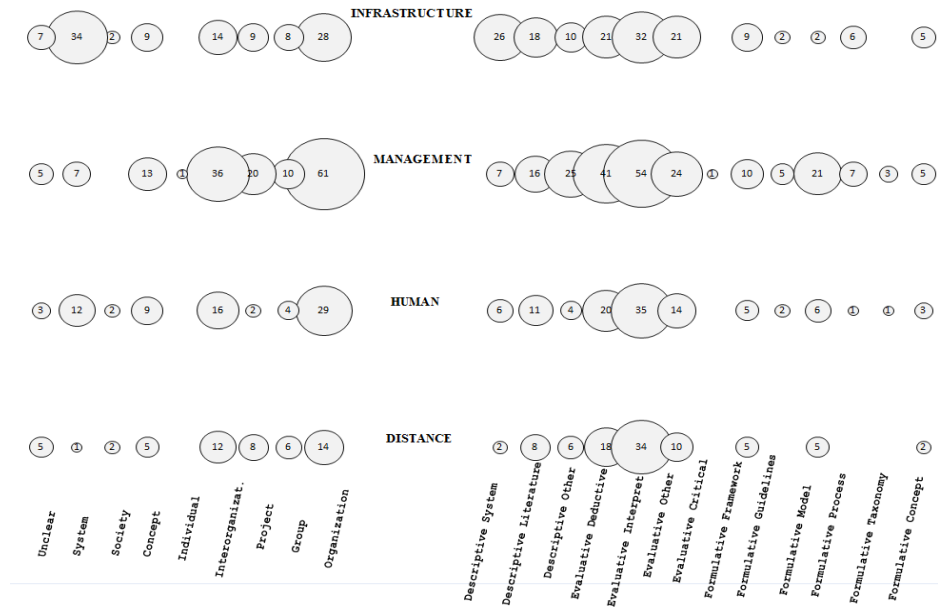

**Fig. 4.** Bubble plot analysis

## 4. THREATS TO VALIDITY

One of the main threats to the validity of our study is the incomplete selection of primary studies or missing relevant studies, even though we followed a systematic process. In order to mitigate this risk we formulated a wide variety of search-terms. These terms were taken from related systematic mapping and systematic literature review (SLR) studies and were updated based upon the retrieved results. Initially, we ensured that at least those SM/SLR studies were indeed retrieved using the search terms drawn from each study. In the next stage, we constructed a sample list of studies from various ICGSE proceedings and ensured that the search terms retrieved these studies as well. During this process the search terms were continuously updated until all sample studies were retrieved, similar to the approach taken by [6]. A second validity threat arises due to researcher bias during the classification process. In order to reduce this threat, we carried out some sample classifications collectively. Furthermore, the lists of studies as classified by the first author were validated by the senior researchers involved. A high level of agreement was achieved, giving us confidence that the classification process was executed appropriately and consistently.

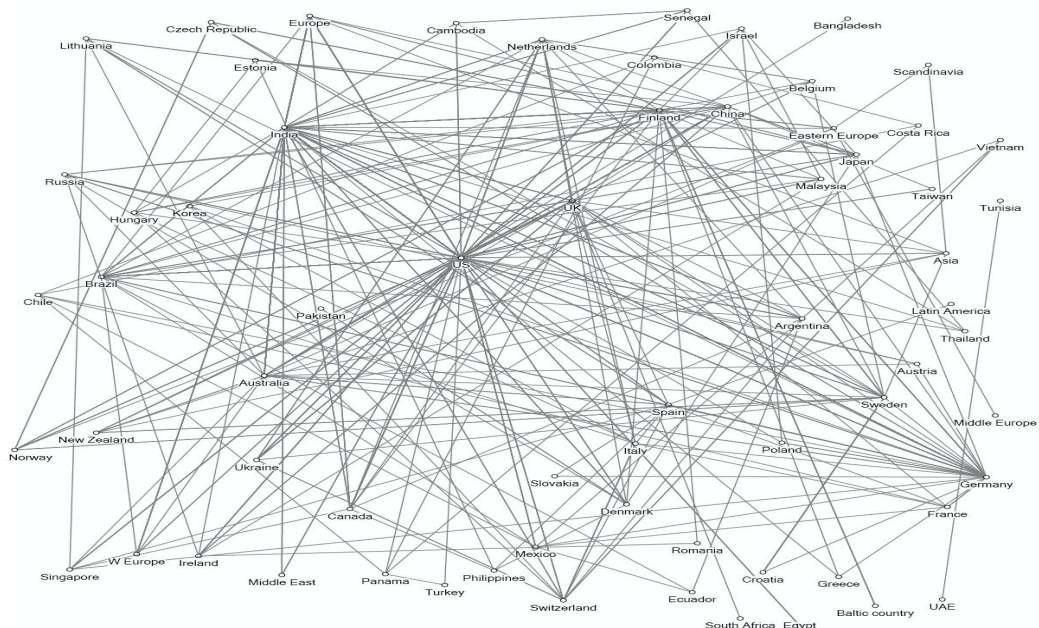

**Fig. 5.** Inter-country relationship analysis

Table 6. Inter-country relationships

| Loc_A | Loc_B | # | Loc_A | Loc_B | # | Loc_A | Loc_B | # | Loc_A | Loc_B | # |
|---|---|---|---|---|---|---|---|---|---|---|---|
| Ind | US | 69 | W.Eu | Ind | 6 | Ita | Che | 3 | Eu | Jpn | 2 |
| Chn | US | 23 | Brazil | Ind | 4 | Jpn | Ind | 3 | Fin | Jpn | 2 |
| Deu | Ind | 15 | Fin | Deu | 4 | Nor | Fin | 3 | Fin | Bra | 2 |
| Bra | US | 11 | Ind | Swe | 4 | Esp | Deu | 3 | Fra | Deu | 2 |
| Aus | US | 10 | Ind | Arg | 4 | US | Che | 3 | Deu | Cze | 2 |
| Eu | US | 10 | Nld | US | 4 | US | Swe | 3 | Ind | Che | 2 |
| Uk | US | 10 | Nld | Ind | 4 | US | Esp | 3 | Ind | M.East | 2 |
| Deu | US | 9 | SGP | US | 4 | US | Sen | 3 | Irl | Chn | 2 |
| Uk | Ind | 8 | US | Ukr | 4 | Chn | Jpn | 3 | Ltu | US | 2 |
|  |  |  | US | SGP | 4 | US | Mex | 3 | Mys | Ind | 2 |
| Fin | Ind | 8 | US | Rus | 4 | US | Mys | 3 | Nld | Ukr | 2 |
| Aus | Ind | 7 | US | Isr | 4 | US | Egy | 3 | Asia | US | 2 |
| Ind | Eu | 7 | US | Nor | 4 | US | Dnk | 3 | Nzl | US | 2 |
| Fin | US | 7 | Ind | Jpn | 4 | Nld | UK | 3 | Nor | Swe | 2 |
| US | Arg | 7 | E.Eu | Fin | 3 | Aus | Esp | 2 | Nor | Cze | 2 |
| US | Ukr | 6 | Fin | Swe | 3 | Aus | Deu | 2 | Esp | Ltu | 2 |
| US | Can | 6 | Fin | Ltu | 3 | Bel | US | 2 | Che | Vnm | 2 |
| Hrv | Swe | 5 | Fin | Baltic. | 3 | Bra | UK | 2 | Che | Ukr | 2 |
| Cze | Fin | 5 | Deu | Rus | 3 | Khm | Senl | 2 | US | Twn | 2 |
| Jpn | US | 5 | Deu | Bra | 3 | Khm | Ind | 2 | US | M.East | 2 |
| Swe | Hrv | 5 | Ind | Sen | 3 | W Eu | US | 2 | US | Khm | 2 |
| US | Jpn | 5 | Dnk | Ind | 3 | Can | Ind | 2 |  |  |  |
| US | Irl | 5 | Ind | Chn | 3 | Can | Eu | 2 |  |  |  |

## 5. CONCLUSIONS

Through this study we have provided a current snapshot of the recent GSE-related research literature. We first classified 275 empirical and non-empirical studies, published between January 2011 and June 2012, into predefined categories (see http://tinyurl.com/GSE-Papers), and we then augmented our analysis with the consideration of the papers published in ICGSE'13. We examined the following characteristics: GSE factors, research approaches, research methods, level of analysis, and GSE project locations. The GSE factors most frequently researched were related to management and infrastructure using evaluative approaches and taking an organizational perspective as the level of analysis. Regarding research methods, interviews, surveys, case studies and field studies are the most commonly used. In relation to project locations, the USA and India are the predominant nations involved in global software projects. Inter-country network analysis also shows that USA-India collaboration is at the top followed by that between the USA and China. It will be interesting to carry out further similar snapshot studies on an on-going basis to see if or how these trends evolve. Similarly, studies could be carried out retrospectively on previous years' research literature to enable comparisons with this study. This study aims to provide a stepping stone for such related studies.

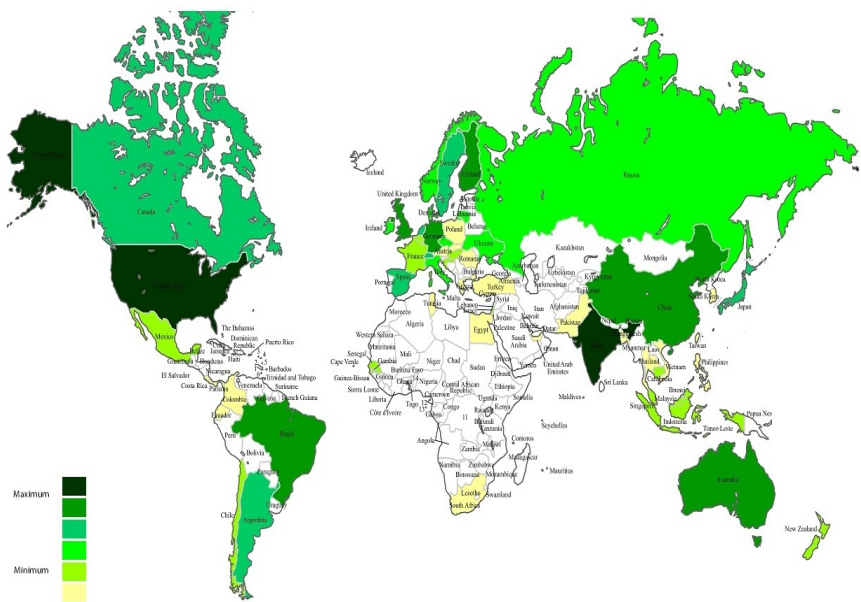

Fig. 6. Locations involved in GSE projects

It appears that, in general, existing solutions are being applied in a GSE context, even though these solutions may lack specific considerations needed for GSE. For instance, aspects of non-functional requirements and stage/phase-related issues are not addressed separately in the current GSE literature. Although the field of GSE research has grown rapidly in terms of the number of studies conducted, these studies are quite narrowly focused towards exploratory research and the provision of explanatory theories. Furthermore, in spite of GSE providing a natural and potentially fruitful setting for critical research, such work is yet to be conducted. The current research focus is mainly directed to organizational concerns, leaving much scope for consideration of the needs of stakeholder groups and individuals. The research is also skewed towards projects having two locations, showing a dearth of studies relating to multiple locations and their underlying complex relationships. Finally, there are regions of the world that are not being currently studied by researchers and it may be useful to con- sider them in the future studies, particularly if the dimensions of culture and their impact on GSE are of interest.

# 6 FUTURE WORK

A notable omission in the current focus of work relating to GSE is any sustained coverage of issues to do with power and exploitation. While the human factors tabulated in Table 2 above include some focus on the factors of fear, trust, cooperation and relationship, these are given relatively limited attention. Again in Figure 4 there is a noted absence of studies at an individual unit of analysis. There are no studies giving personal narratives or biographies – are the workers in GSE deliberately kept invisible? Is this absence a function of the research methods used, for instance, no examples of critical evaluative work have been identified in this review? Or is it an abrogation of our duties as academics to act in the role of 'critic and conscience of society'? Will the future see more equal partnerships in sustainable global ventures, or will there be a backlash against crude models of global labor arbitrage? What risks might that pose to a multi-billion dollar industry? These issues warrant more attention by researchers, although difficult to confront. In addition such research will be challenging to design and conduct, yet the absence of critical evaluative studies presents a glaring gap in current GSE research.

## REFERENCES

1. Šmite, D., Wohlin, C., Gorschek, T., Feldt, R.: Empirical evidence in global software en- gineering: a systematic review. Empirical Software Engineering 15, 91–118 (2009)
2. Petersen, K., Feldt, R., Mujtaba, S., Mattsson, M.: Systematic Mapping Studies in Soft- ware Engineering. In: Proceedings of the 12th International Conference on Evaluation and Assessment in Software Engineering, pp. 68–77 (2008)
3. Bailey, J., Budgen, D., Turner, M., Kitchenham, B., Brereton, P., Linkman, S.: Evidence relating to Object-Oriented software design: A survey. In: First International Symposium on Empirical Software Engineering and Measurement, pp. 482–484 (2007)
4. Fauzi, S.S.M., Bannerman, P.L., Staples, M.: Software Configuration Management in Global Software Development: A Systematic Map. In: Proceedings of the 17th Asia Pacific Software Engineering Conference, pp. 404–413 (2010)
5. Steinmacher, I., Chaves, A.P., Gerosa, M.A.: Awareness Support in Distributed Software Development: A Systematic Review and Mapping of the Literature. Computer Supported Cooperative Work 22, 113–158 (2013)
6. Jalali, S., Wohlin, C.: Agile Practices in Global Software Engineering - A Systematic Map. In: Proceedings of the 5th International Conference on Global Software Engineering, pp. 45–54 (2010)
7. Silva, F.Q.B., Prikladnicki, R., França, A.C.C., Monteiro, C.V.F., Costa, C., Rocha, R.: An evidence-based model of distributed software development project management: results from a systematic mapping study. Journal of Software: Evolution and Process 24, 625–642 (2012)
8. Portillo-Rodríguez, J., Vizcaíno, A., Piattini, M., Beecham, S.: Tools used in Global Soft- ware Engineering: A systematic mapping review. Information and Software Technology 54, 663–685 (2012)
9. Kitchenham, B., Charters, S.: Guidelines for performing Systematic Literature Reviews in Software Engineering. EBSE Technical Report EBSE-2007-01 (2007)
10. Glass, R.L., Vessey, I., Ramesh, V.: Research in software engineering: an analysis of the literature. Information and Software Technology 44, 491–506 (2002)
11. Richardson, I., Casey, V., McCaffery, F., Burton, J., Beecham, S.: A Process Framework for Global Software Engineering Teams. Information and Software Technology 54, 1175–1191 (2012)
12. Dieste, O., Padua, A.G.: Developing Search Strategies for Detecting Relevant Experiments for Systematic Reviews. In: First International Symposium on Empirical Software Engineering and Measurement, pp. 215–224 (2007)
13. Smite, D., Wohlin, C., Feldt, R., Gorschek, T.: Reporting Empirical Research in Global Software Engineering: A Classification Scheme. In: Proceedings of the Third International Conference on Global Software Engineering, pp. 173–181 (2008)
14. Guide to the Software Engineering Body of Knowledge. IEEE Computer Society (2004)
15. Clear, T., MacDonell, S.G.: Understanding technology use in global virtual teams: Re- search methodologies and methods. Information and Software Technology 53, 994–1011 (2011)
16. Myers, M.D., Klein, H.K.: A Set of Principles For Conducting Critical Research In Information Systems. MIS Quarterly 35, 17–36 (2011)
17. Gregor, S.: The nature of theory in information systems. MIS Quarterly 30, 611–642 (2006)
18. Smith, M.A., Shneiderman, B., Milic-Frayling, N., Mendes Rodrigues, E., Barash, V., Dunne, C., Capone, T., Perer, A., Gleave, E.: Analyzing (social media) networks with No- deXL. In: Proceedings of the Fourth International Conference on Communities and Technologies, pp. 255–263 (2009)
19. Dibbern, J., Goles, T., Hirschheim, R., Jayatilaka, B.: Information Systems Outsourcing: A Survey and Analysis of the Literature. ACM SIGMIS Database 35, 6–102 (2004)
20. Butler, N., Slack, F., Walton, J.: IS/IT Backsourcing - A Case of Outsourcing in Reverse? In: Proceedings of the 44th Hawaii International Conference on System Sciences, pp. 1–10 (2011)